\documentclass[prl,twocolumn]{revtex4-1}

\usepackage{graphicx}
\usepackage{color}
\usepackage{amssymb}
\usepackage{bm}
\usepackage{amsmath,amsfonts,latexsym}

\usepackage{braket} 

\usepackage{comment}

\usepackage{bm} 

\newcommand*{\br}{\mathbf{r}}

\newcommand*{\cZ}{{\cal Z}}

\begin{document}

\title{Boson pairing and unusual criticality in a generalized XY model}
\author{Yifei Shi, Austen Lamacraft, and Paul Fendley} 
\affiliation{Department of Physics, University of Virginia,
Charlottesville, VA 22904-4714 USA}
\date{\today}
\date{\today}

\begin{abstract}
We discuss the unusual critical behavior of a generalized XY model
containing both $2\pi$-periodic and $\pi$-periodic couplings between
sites, allowing for ordinary vortices and half-vortices. 
The phase diagram of this system includes both single-particle condensate and pair-condensate phases. 
Using a field theoretic formulation and worm algorithm Monte Carlo simulations, we show that in two dimensions it is possible for the system to pass directly from the disordered (high temperature) phase to the single particle (quasi)-condensate via an Ising transition, a situation reminiscent of the `deconfined criticality' scenario.
\end{abstract}

\maketitle

\begin{figure}
	\centering
		\includegraphics[width=\columnwidth]{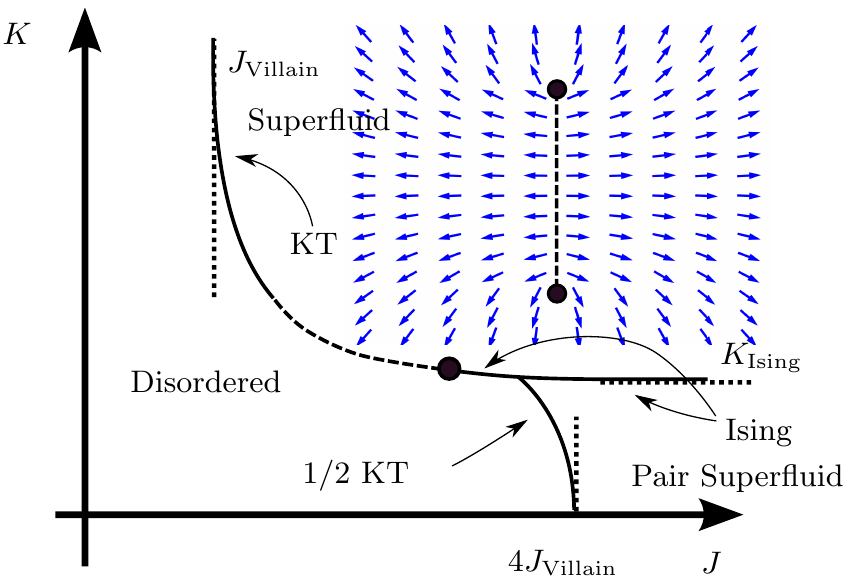}
	\caption{Schematic phase diagram of the model Eq.~\eqref{Partition}. Solid lines are continuous transitions of the type indicated; the dashed line is first order. The dotted line indicates the region where the nature of the transition is unknown. The Renormalization Group analysis developed here applies in the vicinity of the solid dot. Inset: an illustration of a half vortex pair and the associated string.}
	\label{fig:XYphase}
\end{figure}

More than 25 years ago, Korshunov \cite{korshunov:1985} and Lee and Grinstein \cite{lee1985} discussed the statistical mechanics of certain generalizations of the familiar XY model. XY models are the simplest systems capturing the rich physics of vortices, and these topological defects are well known to govern the critical behavior in two dimensions \cite{kosterlitz:1973}. The generalizations discussed by the above authors give rise to half vortices, about which the XY order parameter winds by $\pi$, connected by strings with a finite tension. This in turn leads to a far richer phase diagram (see Fig.~\ref{fig:XYphase}), whose essential elements were verified by numerical simulation \cite{carpenter1989phase}.

Over time there has been considerable interest in identifying systems
where the physics of half vortices and strings plays a role, ranging
from nematic liquid crystals \cite{lee1985,geng:2009} to the A-phase
of $^{3}\text{He}$ \cite{korshunov:1985,korshunov2006}, to spinor Bose
condensates \cite{mukerjee2006,james2011}. Indeed, half vortices have
now been directly observed in exciton-polariton condensates
\cite{lagoudakis2009}. In recent years a large amount of activity has
focused on one particular candidate: a gas of attractive bosons. It
can have two distinct superfluid phases: an atomic superfluid of
bosons and a molecular superfluid of boson pairs, with the latter
supporting half vortices \cite{romans:2004,radzihovsky:2004}. Such a
system is in general likely to be unstable to collapse, but a possible
resolution of this difficulty is to harness three-body loss to project
out triple occupancy of each site of an optical
lattice \cite{daley:2009}. This idea led to a resurgence of interest
in the problem \cite{diehl:2010,lee:2010,bonnes:2011,ng:2011,ejima:2011}.

We have looked anew at the phase diagram of this
type of system. Previous studies found that one could either pass
directly from the normal state to the atomic condensate at low
temperatures, or first into a molecular condensate with no
single-particle long-range order, and thence to the atomic condensate
in an Ising transition. Remarkably, we find that in two dimensions
there is a region of the finite temperature phase diagram where the normal-to-atomic
superfluid transition is of the Ising, rather than the
Kosterlitz--Thouless type (see Fig.~\ref{fig:XYphase}). Such
unconventional behavior is reminiscent of the `deconfined criticality'
scenario \cite{senthil2004}. The two have a common origin in that the
expected proliferation of point-like defects is suppressed by
critical fluctuations. Our prediction is further supported by Monte Carlo simulations using the worm algorithm and a novel method based on conformal field theory to identify an Ising transition in a system with non-Ising degrees of freedom. We stress that our result applies to the finite temperature phase diagram of many of the above mentioned systems in 2D (see e.g. \cite{romans:2004,radzihovsky:2004,geng:2009,james2011}), as well as to the zero temperature phase diagram of the 1D system in Ref.~\cite{ejima:2011}.

The simplest generalization of the XY model with the requisite physics has the form
\begin{equation}
	\label{genXYlett_genXYham}
	H = -\sum_{\langle ij\rangle} \left[(1-\Delta)\cos(\theta_{i}-\theta_{j})+\Delta\cos\left(2\theta_{i}-2\theta_{j}\right)\right]
\end{equation}
where the angular variables $\theta_{i}:0\leq \theta_{i}<2\pi$ are defined on a square lattice and $\langle ij\rangle$ represents nearest neighbor pairs. In quantum mechanical language, the two terms correspond to the hopping of single bosons and boson pairs respectively.
$\Delta=0$ recovers the usual XY model, which allow for the existence of point-like vortices around which the phase increases by a multiple of $2\pi$. $\Delta=1$ corresponds to an XY model with half the periodicity, and thus half vortices of $\pi$ winding, but otherwise identical properties. For $\Delta>1/5$ the intersite energy develops a metastable minimum at $|\theta_{i}-\theta_{j}|=\pi$. Pairs of half vortices are connected by a `string' where $\theta_{i}-\theta_{j}$ jumps by $\pi$ and for $\Delta<1$ this string has a finite tension (see Fig.~\ref{fig:XYphase}).

When this tension is large (small $\Delta$) half vortices are bound
together irrespective of their topological charge, so that only
integer vortices may exist freely. Thus here the model displays the familiar Kosterlitz--Thouless (KT) transition. When the tension is small ($\Delta$ close to 1), a KT transition of the half vortices can occur ($\frac{1}{2}$KT). After this transition half vortices are bound together in pairs, so that the strings connecting them form closed domain walls, which disappear at a lower temperature as the tension overcomes their entropy. This second transition is of the Ising type.

The resemblance of strings to Ising domain walls that can terminate on
half vortices suggests that the operator inserting a half vortex
includes an Ising disorder operator \cite{kadanoff1971}. In the high
temperature phases (corresponding to the disordered and pair
superfluid phases in Fig.~\ref{fig:XYphase}) the strings are not
confining, equivalent to long range order for the disorder
operator. The half vortices then can drive a KT transition by the
familiar mechanism. However, along the Ising critical line the
expectation value of the disorder operator vanishes. The resulting
critical fluctuations of the strings suppress the proliferation of
half vortices, so that a direct Ising transition between the
disordered and superfluid phases can occur.

\emph{Villain model}. 
This qualitative picture is borne out by our analysis of a particular
microscopic model. Following the original references
\cite{korshunov:1985,lee1985}, we study a variant of the model Eq.~\eqref{genXYlett_genXYham} with partition function
\begin{equation}
	\label{Partition}
	\cZ =
	\prod_{c}\int_{-\pi}^{\pi}\frac{d\theta_{c}}{2\pi}\prod_{\langle
	ab\rangle} 
\left(w_{V}(\theta_{ab})+e^{-K}w_{V}(\theta_{ab}-\pi)\right)
\end{equation}
where $\theta_{ab}\equiv \theta_a-\theta_b$ and the Villain potential
$w_V$ is
\begin{equation}
	\label{Villain}
	w_{V}(\theta)\equiv \sum_{p=-\infty}^{\infty}e^{-\frac{J}{2}\left(\theta+2\pi p\right)^{2}}\propto\sum_{j=-\infty}^{\infty}e^{in\theta} e^{-\frac{J_{*}}{2}n^{2}}
\end{equation}
with $J_{*}=J^{-1}$. $K\to\infty$ is the usual Villain model, with a KT transition at $J_{\text{Villain}}\sim 0.75\ldots$ \cite{hasenbusch:1994,hasenbusch:1997}, while $K=0$ corresponds to a $\pi$ periodic Villain model, allowing free half vortices, with a $\frac{1}{2}$KT transition at $J=4J_{\text{Villain}}\sim 3$. Finite $K$ gives the strings connecting half vortices a finite domain wall energy. When $J\to \infty$ the phase differences between neighboring sites are restricted to $0$ or $\pi$ and the usual square lattice Ising model with transition at $K_{\text{Ising}}=\ln(1+\sqrt{2})\sim 0.881\ldots$ is recovered. 

We map to a generalized height model using the second representation
of the Villain potential in Eq.~\eqref{Villain}. Integrating out the
angular variables leaves
\begin{equation}
	\label{XYVillain_Zintegrated}
	\cZ=\sum_{\substack{\left\{n_{ij}\right\}\\ \nabla\cdot n=0}}\exp\left(-\frac{J_{*}}{2}\sum_{\langle ij\rangle} n_{ij}^{2}+\frac{K_{*}}{2}\sum_{\langle ij\rangle}(-1)^{n_{ij}}\right),
\end{equation}
where $K_{*}$ and $K$ satisfy the Kramers--Wannier duality relation	$\sinh K_{*}\sinh K =1$. Here the variables $n_{ij}$ live on the links of the lattice. Due to their vanishing lattice divergence, they may be thought of as currents describing the world lines of bosons. 


Because $\nabla\cdot n_{ij}=0$, we may write $n_{ij}=h_{i}-h_{j}$,
with integer-valued heights $\left\{h_{i}\right\}$ on the dual
lattice. Then 
%
%
\begin{equation}
	\label{XYVillain_height}
	\cZ=\sum_{\{h_{i}\}}\exp\left(-\frac{J_{*}}{2}\sum_{\langle ij\rangle} (h_{i}-h_{j})^{2}+\frac{K_{*}}{2}\sum_{\langle ij\rangle}\mu_{i}\mu_{j}\right),
\end{equation}
where the variables $\mu_{i}\equiv (-1)^{h_{i}}$. The discrete Gaussian model given by the first term of Eq.~\eqref{XYVillain_height} has a roughening transition from a smooth phase at small $J$ (high temperature) to a rough phase at large $J$, corresponding to a superfluid state of the bosons. The new term in $K_{*}$ allows for the existence of a second rough phase where the heights are predominantly even or odd, with the currents $n_{ij}$ even, corresponding to a pair superfluid.

The Ising and Gaussian parts are disentangled by writing
$h_{i}=2\tilde h_{i}+(\mu_{i}-1)/2$ and identifying the $\mu_i$ as the disorder
variables of an Ising model, dual to the usual spin variables. We move to continuous $\tilde h_{i}$ by introducing the $\delta$-functions $\prod_{i}\sum_{p}\delta(\tilde h_{i}-p)=\prod_{i}\sum_{q}e^{2\pi i q \tilde h_{i}}$. An effective sine--Gordon description is arrived at in the standard manner \cite{kogut1979} by introducing an effective coupling (vortex fugacity) $z_{p}$ for each harmonic. After a shift in the integration variables $\tilde h_{i}\to \tilde h_{i}-(\mu_i+1)/4$ and retaining only the $q=\pm 1,\pm 2$ terms we arrive at
\begin{equation*}
\begin{split}
	\cZ&=\sum_{\mu_{i}=\pm 1}\int \prod_{i} d \tilde{h}_{i}
		\exp\left(\sum_{\langle ij\rangle} S_{ij} + \sum_i V_i\right)\ ,\\
		\label{XYVillain_continuous_shift}
	S_{ij}&=-2J_{*}\left(\tilde{h}_{i}-\tilde{h}_{j}\right)^{2}
	+\frac{K_{*}}{2}\mu_{i}\mu_{j}\ ,\\
	\nonumber
	V_{i}&=z_{1}\,\mu_{i}\cos(2\pi \tilde{h}_{i}) 
	+z_{2}\cos(4\pi \tilde{h}_{i})\ .	
\end{split}	
\end{equation*}
The nearest-neighbor interactions $S_{ij}$ are those
of decoupled discrete Gaussian and Ising models. The two
terms in the potential $V_i$ describe respectively the half and
integer vortices, making precise how the half vortices couple the
Ising and Gaussian degrees of freedom.

%
%
%
%
%

\emph{Renormalization Group (RG) analysis}. The limits shown in
Fig.~\ref{fig:XYphase} provide the skeleton of the phase diagram for
this model, but to understand what happens when the transitions
approach each other requires an analysis of the coupling between the
Gaussian and the Ising degrees of freedom. Since we are concerned principally with the \emph{nature} of the transition in this region, this is most conveniently
accomplished in the field theory limit, where we have a sine-Gordon model and an Ising model coupled by the perturbation
\begin{equation}
	\label{genXYlett_IsingGaussian}
	H_{\frac{1}{2} \text{-V}}=z_{1}\int d\mathbf{x}\,
	\mu(\mathbf{x})\cos[2\pi \tilde h(\mathbf{x})]\ .
\end{equation}
In the high temperature (low $K$) phase of the Ising model, the
disorder operator $\mu(\br)$ acquires an expectation value. Here
$H_{\frac{1}{2}\text{-V}}$ becomes equivalent to the cosine potential of
the sine-Gordon model. When $z_1=z_2=0$, the operator $\cos(2\pi
n\tilde{h})$ has dimension $\pi n^2/(4J_{*})$, and so is relevant for
$J<8/(\pi n^2)$. Thus in the Ising high-temperature phase, the
half-vortices and vortices drive KT transitions at $J=8/\pi$ 
and $J=2/\pi$ respectively. (These should
be understood as the \emph{renormalized} values of $J$: the
transitions occurs at a larger value of \emph{bare} J.) This
accounts for the fourfold increase in the jump of the superfluid
density at the $\frac{1}{2}$KT transition relative to the usual one
 \cite{korshunov1986,mukerjee2006}.
%
%

The unusual behavior occurs along the critical line of the Ising
model. Here $\mu(\mathbf{x})$ no longer has an expectation value, but only
critical fluctuations obeying $\langle\mu(\mathbf{x})\mu(\mathbf{y})\rangle\sim |\mathbf{x}-\mathbf{y}|^{-1/4}$. The scaling dimension of $\mu(\mathbf{x})\cos[2\pi \tilde h(\mathbf{x})]$ is $\frac{1}{8}+\frac{\pi}{4J_{*}}$, which does not become relevant until $J=\frac{15}{2\pi}$, a smaller value than for proliferation of half vortices in the disordered phase. This suggests the scenario depicted in Fig.~\ref{fig:XYphase}: the Ising transition persists \emph{after} it has met the $\frac{1}{2}$KT transition.

\begin{figure}
	\centering
		\includegraphics[width=0.49\columnwidth]{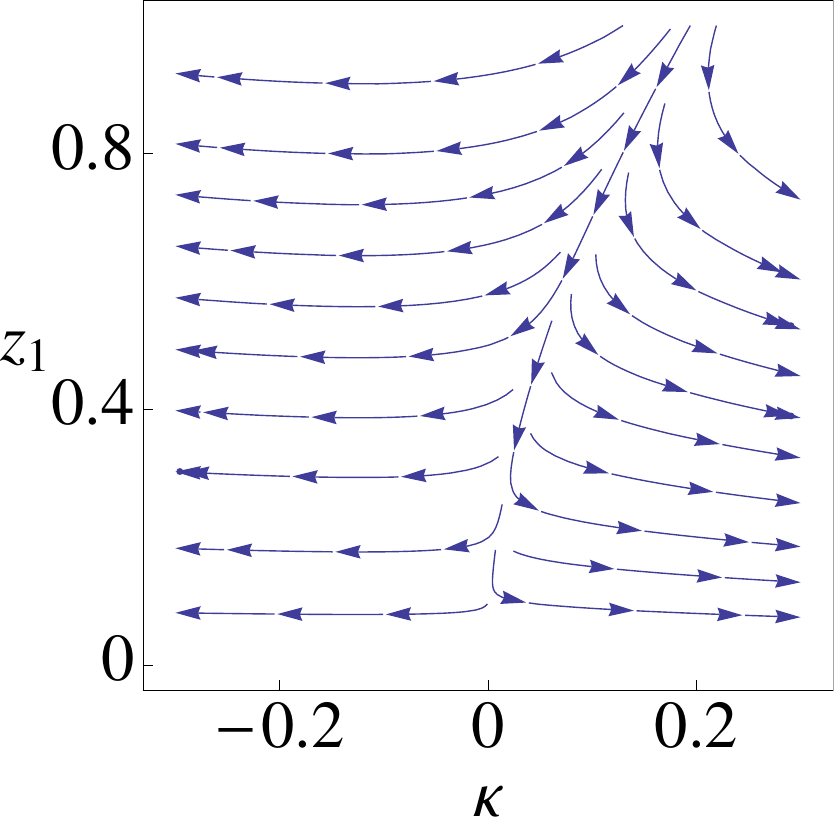}
		\includegraphics[width=0.49\columnwidth]{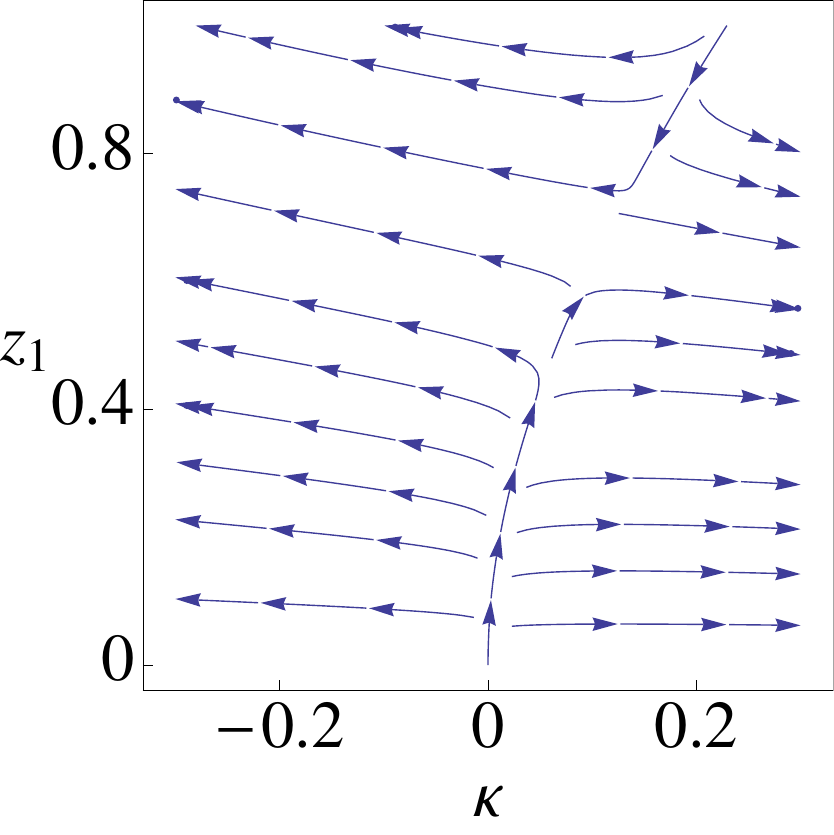}
	\caption{(Left) RG flow in the $\kappa-z_{1}$ plane for $J=\frac{31 }{4\pi}$. Here the $\mu \cos(2\pi \tilde h)$ perturbation is irrelevant, even though $\langle \mu\rangle\cos(2\pi \tilde h)$ is relevant. Note that the Ising fixed point at $\kappa=0$, $z_{1}=0$ is stable. (Right) RG flow for $J=\frac{29}{4\pi}$. The Ising fixed point is now unstable.}
	\label{fig:RGFlow}
\end{figure}

The RG equations give more insight into these transitions. 
Defining the deviation from Ising criticality to be $\kappa=K-K_{c}$,
at $z_2=0$ we find to second order in $\kappa$ and $z_{1}$ \cite{cardy:1996}
\begin{equation}\label{XYVillain_RGeq}
\begin{split}
	\frac{dz_{1}}{dl}=\left(\frac{15}{8}-\frac{\pi}{4J_{*}}\right)z_{1}-\frac{\kappa z_{1}}{2},\\
	\frac{dJ_{*}}{dl}=\frac{\pi^{2}z_{1}^{2}}{4}, \qquad
	\frac{d\kappa}{dl}=\kappa-\frac{z_{1}^{2}}{4}\ ,	
\end{split}
\end{equation}
where $l=\log(\xi/\xi_{0})$, with $\xi$ being the coarse-grained length scale. In Fig.~\ref{fig:RGFlow} we show two sections of the flow in the $\kappa-z_{1}$ plane, for values of $J$ above and below $J_{c}\equiv\frac{15}{2\pi}$. When $J>J_{c}$, the Ising fixed point at the origin is stable to the $H_{\frac{1}{2}\text{-V}}$ perturbation. Note, however, that this is a \emph{dangerously irrelevant} perturbation, with $z_{1}$ growing at negative $\kappa$ (higher temperature) reflecting the proliferation of the half vortices. In contrast, when $J<J_{c}$, the Ising fixed point is unstable. The apparent fixed point at finite $z_{1}$ is in fact a separatrix along which $J$ flows to zero: crossing this separatrix presumably corresponds to a first order transition. 

The RG equations Eqs.~\eqref{XYVillain_RGeq} give are valid in the vicinity of the Ising fixed point. In particular, our conclusion regarding the dangerous irrelevance of $H_{\frac{1}{2}\text{-V}}$ applies close to the critical value $J_{c}=\frac{15}{2\pi}$ where this perturbation becomes marginal. When $J>\frac{8}{\pi}$ we know that $H_{\frac{1}{2}\text{-V}}$ is irrelevant even deep in the disordered phase.

\begin{figure}
	\centering
		\includegraphics[width=\columnwidth]{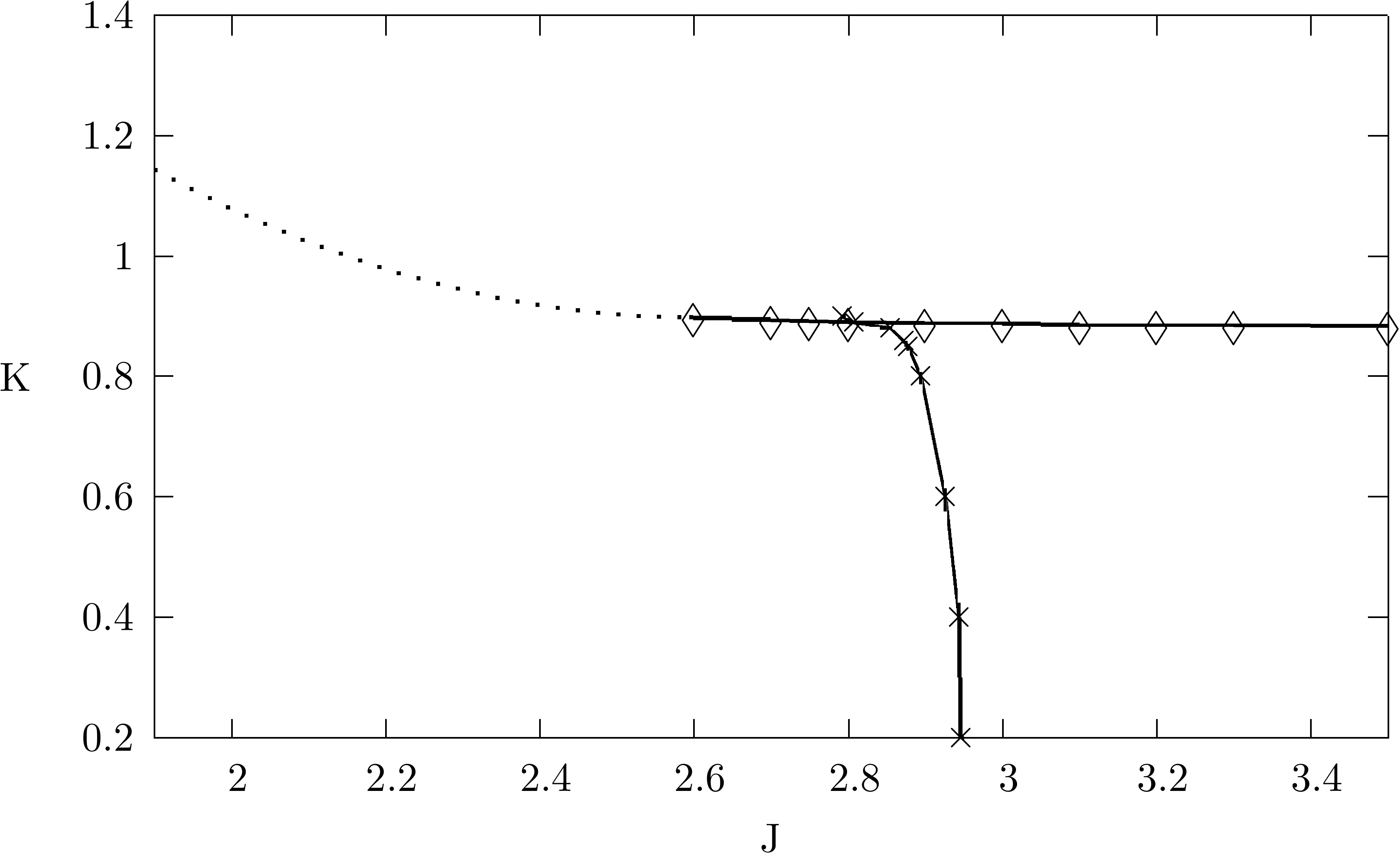}
	\caption{Phase diagram near the meeting of the Ising ($\Diamond$) and $\frac{1}{2}KT$ ($\times$) transitions determined by numerical simulation. The transitions are determined as described in the text with the Ising line stopping at the last point where we are confident of the nature of the transition. The dotted line is a sketch of the (presumably) first order line connecting to the usual KT transition (see Fig.~\ref{fig:XYphase}).}
	\label{fig:Simulations_Data_PhaseDiagramNearTricriticalPoint}
\end{figure}

\emph{Numerical simulations}. We have tested the above calculations
using Monte Carlo simulations of the model Eq.~\eqref{Partition} based
on the worm algorithm \cite{prokofev:2001}, in the formulation given
in Ref.~\cite{alet2010}.  Both single and double worms are needed to
accurately simulate the paired phase. The worm algorithm directly
simulates the partition sum in term of currents,
Eq.~\eqref{XYVillain_Zintegrated}.  The phase diagram resulting from
our simulations is illustrated in
Fig.~\ref{fig:Simulations_Data_PhaseDiagramNearTricriticalPoint},
clearly showing the persistence of the Ising transition after the
Ising line meets the $\frac{1}{2}$KT line. 

A particular advantage of the worm algorithm is that it simultaneously
simulates all \emph{sectors} of the model defined with periodic
boundary conditions (i.e. on a torus). A sector is specified by a pair
of integers $\mathbf{W}=(W_x,W_y)$ giving the winding of the
current loops around each circle of the torus; in terms of heights,
$h(x+L,y)= h(x,y) + W_{y}$ and $h(x,y+L)= h(x,y) + W_{x}$.  To locate
the KT and $\frac{1}{2}$KT phase boundaries, we follow the method of
Ref.~\cite{harada1998kosterlitz} and exploit the
fact\cite{pollock:1987} that the superfluid density (helicity modulus)
is $\Upsilon =
\frac{T}{2} \langle \mathbf{W}^{2} \rangle$
(we restore temperature, which provides the energy scale). 

%
%

\begin{figure}
	\centering
		\includegraphics[width=\columnwidth]{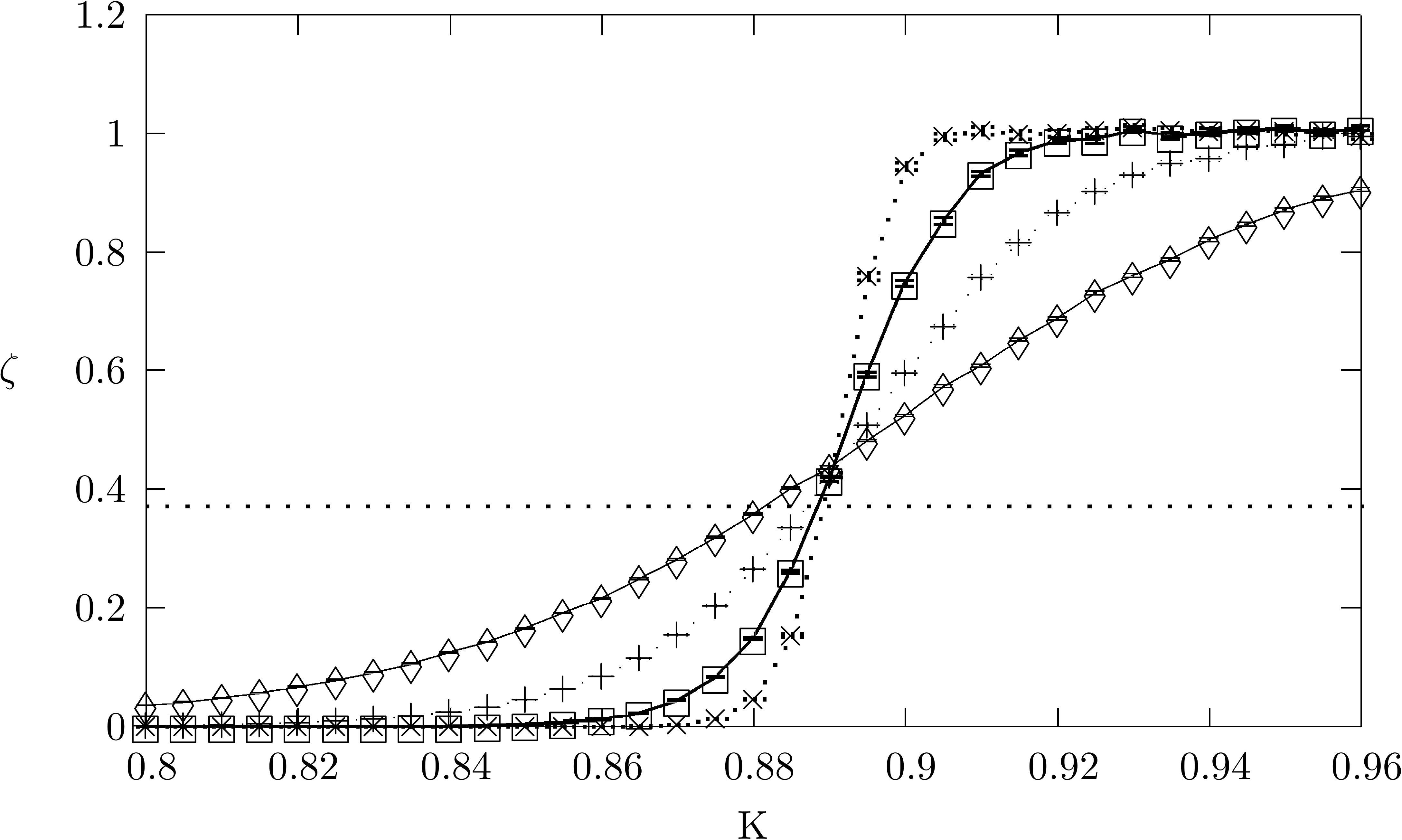}
	\caption{The ratio Eq.~\eqref{genXYlett_ratio} of partition functions is used to locate the Ising transition from the crossing point for different system sizes. Here $\Diamond$, $+$, $\square$, $\times$ symbols are for systems sizes $L=20,40,80,160$ respectively, $J=2.8$, and the (small) error bars are included. The dotted line corresponds to the pure Ising value $\zeta_c=.3729$.}
	\label{fig:Crossing}
\end{figure}

\begin{figure}
	\centering
		\includegraphics[width=\columnwidth]{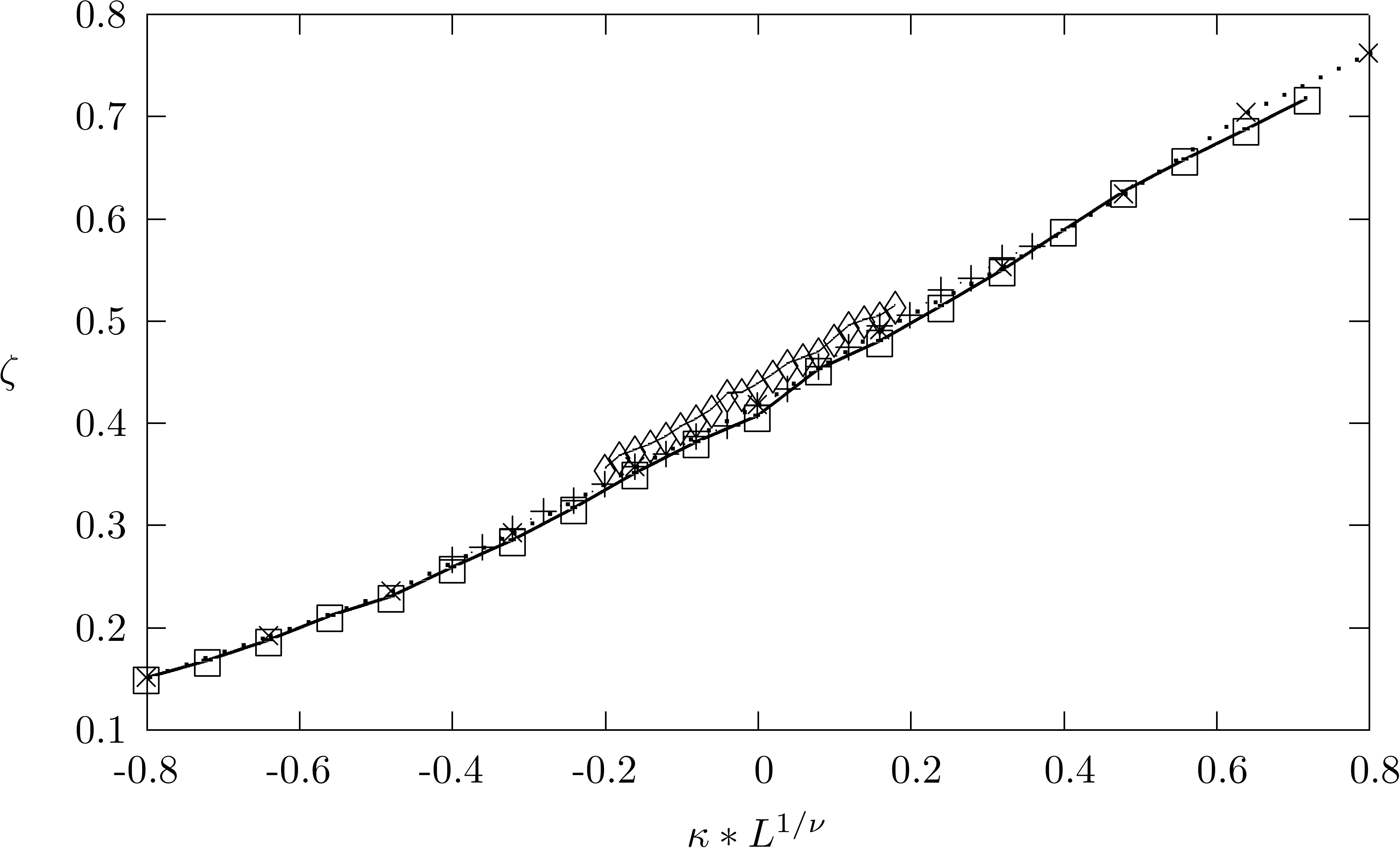}
	\caption{$\zeta$ plotted \emph{vs.} $\kappa L^{1/\nu}$. For the same $L$ as in Fig.~\ref{fig:Crossing}, the best collapse is achieved for $\nu=1.00\pm 0.02$}
	\label{fig:RescaledJ=28}
\end{figure}

To locate the Ising transition, we utilize
the partition functions in the various sectors.
The non-universal bulk contribution to the free
energy is independent of sector, and so the ratio of partition
functions in different sectors should be a universal property of the
critical point. Thus
\begin{equation}
	\label{genXYlett_ratio}
	\zeta\equiv\frac{\cZ_{W_{x}\text{ odd}, W_{y}\text{ even}}+\cZ_{W_{x}\text{ even}, W_{y}\text{ odd}}}{2\cZ_{W_{x}\text{ even}, W_{y}\text{ even}}}
\end{equation} 
becomes independent of system size, and so the curves for different
sizes plotted in Fig.~\ref{fig:Crossing} cross at the transition.
We extract the correlation length critical exponent
by assuming the form $\zeta(L,J,K)=f(L/\xi)$ with the correlation
length $\xi\sim \kappa^{-\nu}$ applying near the critical point. The
best scaling collapse (Fig.~\ref{fig:RescaledJ=28}) is achieved with
$\nu=1.00\pm 0.02$, consistent with $\nu=1$ for the Ising model.

In fact, the critical ratios themselves are known exactly from
conformal field theory \cite{ginsparg:1988}. Even though the Gaussian
and Ising actions decouple at $z_{1}=0$, the fields still
satisfy a nontrivial `gluing condition'. If $W_{x}$ (say) is odd, then
the $\mu_{i}$ satisfy antiperiodic boundary conditions in the $y$
direction, and $\tilde h(x,y+L)-\tilde h(x,y)$ is half integer. In the
field theory limit, we have at the decoupling point
\begin{equation}
	\label{genXYlett_oddeven}
	\cZ_{W_{x}\text{ odd}, W_{y}\text{ even}}=\cZ^{\text{Ising}}_{PA}\cZ^{\text{Gaussian}}_{0,1/2},
\end{equation}
where `P' (`A') denotes the (anti-)periodic boundary conditions in the
Ising part, and $0$ ($1/2$) denotes (half) integer winding in the
bosonic part.  For the Ising model on a
square torus, 
$\cZ^{\text{Ising}}_{PP}/\cZ^{\text{Ising}}_{AP}=
1+2\vartheta_2(0,e^{-\pi})/\vartheta_3(0,e^{-\pi})=2.68179\dots\ ,$
where the $\vartheta_j(z,q)$ are the standard Jacobi theta functions.
%
%
The Gaussian contribution to $\zeta$ is $\vartheta_{3}(-\pi/2,e^{-\pi^{2}J/2})/\vartheta_{3}(0,e^{-\pi^{2}J/2})$, equal to unity to the fifth significant
digit at $J=J_c$, and increasing to one in the region of interest
$J<J_c$. Thus the ratio $\zeta$ at the Ising critical point is close to
$\zeta_c=.3729$. In Fig.~\ref{fig:Crossing}
the values of $\zeta$ at the crossings are indeed approaching $\zeta_c$ with increasing system size.





In conclusion, we have shown that a two dimensional XY system supporting half vortices and strings has unusual critical behavior driven by the interplay of these two types of defects. The same physics is expected to play a role in the $(1+1)$-dimensional quantum problem (see Refs.~\cite{daley:2009,ejima:2011}), where there is the additional freedom of lattice filling to explore. In three dimensions (or the $2+1$-dimensional quantum problem) the analog of the coupling between the half vortices and the disorder variables of the Ising model is an Ising gauge charge carried by half vortex lines. The consequences of this coupling will be explored in future work.

Our thanks are due to Andrew James, Dan Podolsky, Leo Radzihovsky, and Boris Svistunov for helpful discussions, and to the University of Virginia Alliance for Computational Science and Engineering, especially Katherine Holcomb, for their assistance. The support of the NSF through awards DMR-0846788 (AL) and DMR/MPS-0704666 and DMR/MPS1006549 (PF) and the Research Corporation through a Cottrell Scholar award (AL) is gratefully acknowledged.


%

\end{document}